\pdfoutput=1

\documentclass[12pt,preprint]{aastex}

\slugcomment{Accepted for Publication in ApJ, Nov. 1, 2010}

\shorttitle{AGN Powered by Spin or Accretion}
\shortauthors{McNamara et al.}

\begin{document}
\def\gae{\mathrel{\hbox{\rlap{\hbox{\lower2pt\hbox{$\sim$}}}\hbox{\raise2pt\hbox{$>$}}}}}
\def\lae{\mathrel{\hbox{\rlap{\hbox{\lower2pt\hbox{$\sim$}}}\hbox{\raise2pt\hbox{$<$}}}}}
\def\msunyr{{\rm\, M_\odot\, yr^{-1}}}


\title{Are Radio AGN Powered by Accretion or Black Hole Spin?}


\author{B. R. McNamara$^{1,2,3}$}
\author{Mina Rohanizadegan$^1$}
\author{P. E. J. Nulsen$^3$}
\affil{$^1$University of Waterloo, Department of Physics and Astronomy, Waterloo, Canada\\$^2$Perimeter Institute for Theoretical Physics, Waterloo, Canada\\$^3$Harvard-Smithsonian center for Astrophysics}







\begin{abstract}
We compare accretion and black hole spin as potential energy sources for outbursts from AGN in brightest cluster galaxies (BCGs).
Based on our adopted spin model, we find that the distribution of AGN power estimated from X-ray cavities is consistent with a broad range of both spin parameter and accretion rate.    Sufficient quantities of molecular gas are available in most BCGs to power their AGN by accretion alone.  However, we find no correlation between AGN power and molecular gas mass  over the range of jet power considered here. For a given AGN power, the BCG's  gas mass and accretion efficiency, defined as the fraction of the available cold molecular gas that is required to power the AGN,  both vary by more than two orders of magnitude.  Most of the molecular gas in
BCGs is apparently consumed by star formation or is driven out of the nucleus by the
AGN before it reaches the nuclear black hole.  Bondi accretion from hot atmospheres is generally unable to fuel powerful AGN, unless their
black holes are more massive than their bulge luminosities imply.
We identify several powerful AGN that reside in relatively gas-poor galaxies, indicating an unusually efficient mode of accretion, or that their AGN are powered
by another mechanism.  If these systems are powered primarily by black hole spin, rather than by accretion,  spin must also be tapped efficiently in some systems,  i.e., $P_{\rm jet} > \dot Mc^2$, or their black hole masses must be substantially larger than the values implied by their bulge luminosities. 
We constrain the
(model dependent) accretion rate at the transition from radiatively inefficient to radiatively efficient accretion flows to be a few percent of the Eddington rate, a value that is consistent with other estimates.  
\end{abstract}


\keywords{galaxies: clusters: general -- galaxies: cooling flows -- Active Galactic Nuclei}

\section{Introduction}
\label{sec:intro}

A growing body of evidence suggests that energetic feedback from active galactic nuclei (AGN) is suppressing the   
cooling of hot halos in galaxies and clusters, and preventing significant star formation in bulges at 
late times (B{\^i}rzan et al. 2004, Bower et al. 2006, Best et al. 2006).
But how AGN are powered and how feedback operates are both poorly understood.   AGN are thought to be powered primarily by the gravitational binding energy released from accretion onto nuclear, supermassive black holes (SMBHs).   Much of the accretion energy is released promptly in the form of radiation and mechanical outflows.   However, the accreted angular momentum can spin-up SMBHs (eg., Volonteri et al. 2005), storing rotational 
energy that may be tapped over longer timescales.  Spin may therefore play an important role in the formation of extragalactic radio sources (Begelman, Blandford, \& Rees 1984, Wilson \& Colbert 1995).

There are good reasons to investigate spin-powered feedback in galaxies and clusters. 
First, the energy available in a spinning black hole is significant with respect to the
thermal energy of its X-ray atmosphere.  A  $10^9 M_\odot$  black hole spinning near its maximal rate stores $\gae 10^{62}~\rm erg$ of energy 
which may be released in the form of mechanically-dominated  jets.  X-ray images have shown that jet energy couples efficiently to 
hot atmospheres (B{\^i}rzan et al. 2004, Merloni \& Heinz 2008), which
elevates the entropy (energy) of the hot gas and suppresses cooling and star formation in galactic bulges at late times (Voit \& Donahue 2005).   Second, 
in most spin models,  the jetted outflow is coupled to the black hole's rotational energy through poloidal magnetic fields anchored to an accretion disk
(Blandford \& Znajek 1977, Meier 1999, Beckwith, et al. 2009, Krolik \& Hawley 2010).  
These models generally require the field to be confined by the accreting gas, placing an upper limit on the magnetic field strength and hence the power that can be tapped from spin (i.e., $\dot M \propto B^2_p\propto P_{jet}$).  
The connection between jet power, spin, and accretion rate could in principle provide the physical basis for a feedback loop  (McNamara et al. 2009).
Finally, systems with high jet power and low fuel reserves may have difficulty powering their jets by accretion, 
making spin power an appealing alternative to accretion power.

Radio jets are thought to form in radiatively inefficient accretion flows (e.g., RIAFs or ADAFs) associated with hot, 
thick disks accreting far below the  Eddington accretion rate (Rees et al. 1982, Narayan \& Yi 1995, Narayan \& Quataert 2005, Wu \& Cao 2008).
Several models have been proposed to power and collimate radio jets (Begelman, Blandford \& Rees 1984),
including the  Blandford-Znajek (BZ) mechanism (Blandford and Znajek 1977, Ghosh \& Abramowicz 1997) and its variants. The BZ mechanism derives power from a spinning
black hole through torques applied by magnetic field lines threading the ergosphere and inner region of the accretion flow.  Field lines wind
up along the hole's spin axis creating a collimated outflow in the form of a jet.  Variants of the BZ model, so-called hybrid models, are able to boost the power output per gram of accreted mass through the amplification of magnetic flux in the plunge region of the black hole (Reynolds et al.  2006, Garofalo, Evans, \& Sambruna 2010),
and through frame-dragging (Meier 2001).  

In this paper,  we evaluate the roles of spin and accretion in generating powerful AGN outbursts in the cores of clusters.
Because current jet models require a substantial level of accretion in order to extract spin power (Nemmen et al. 2007, Cao \& Rawlings 2004),
we assume at the outset that all systems are ADAFs accreting
at the same fraction of the Eddington accretion rate, and we then critique this assumption.  Our analysis differs from other analyses in that we do not rely on radio synchrotron
power as a measure of jet power (e.g., Cao \& Rawlings 2004).  Instead, we derive jet power from cluster X-ray cavities taken from the Rafferty et al. (2006) sample, 
which provide reliable mechanical power estimates that can be compared directly to spin models.  
Black hole masses were estimated using R-band absolute magnitudes also taken from Rafferty et al. (2006), and folded through the black hole mass versus bulge magnitude relation of Lauer et al. (2007).   The objects considered here and their properties are listed in Table 1.

\section{Power from a Rotating Black Hole}

We adopt the model of Nemmen et al. (2007) for relating jet power to the parameters of accreting black holes.  Their model follows the hybrid model proposed by Meier (1999, 2001), which relies on the BZ mechanism in rapidly spinning black holes and the Blandford-Payne model (Blandford \& Payne 1982) at lower spins.  Under the BZ mechanism, magnetic field threading the inner edge of the accretion disk and the ergosphere can tap the spin energy of the black hole to power a jet.  The available power is proportional to the square of the spin parameter, $j$.  The power is small unless the spin parameter is close to its maximum value of unity.  Under the Blandford-Payne mechanism, magnetic field threading the inner edge of the disk taps the kinetic energy of disk material to drive the jet.  This mechanism can power jets even for $j = 0$, but the BZ mechanism can tap substantially greater jet powers from rapidly spinning black holes

For a given black hole mass and spin parameter,  the jet power depends quadratically on the poloidal magnetic field strength
(magnetic pressure).
Hybrid spin models (eg., Meier 2001, Nemmen et al. 2007, Benson \& Babul 2009) are able to enhance jet power by
including field amplification from the rotation of the accretion disk, frame dragging, and so forth.
The poloidal magnetic field strength is not an arbitrary parameter,  but is instead determined by the accretion rate through the disk and ergosphere such that $B^2_{P}\propto \dot M$ (Meier 1999, 2001, Nemmen et al. 2007).  
Observational estimates of the spin parameters of
radio galaxies (eg., Cao \& Rawlings 2004, Daly 2009) cannot be decoupled from the
accretion rate, which is an issue we focus on here.

\section{AGN Cavity Power as a Measure of Spin Power}

In Fig. 1 we plot jet power against the estimated black hole mass for the hosting BCGs.  The jet
powers span the range $10^{42}~\rm erg~s^{-1}$ to $10^{46.5}~\rm erg~s^{-1}$, and the black hole masses lie between
$\sim 10^{8.5}~\rm M_\odot$ and $10^{10}~\rm M_\odot$. Four objects with dynamically determined black hole masses are highlighted in blue.   All fall within the range of black hole masses estimated from bulge luminosities.

The total energy associated with a maximally spinning, $10^9 \rm M_{\odot}$ black hole corresponds to $ \sim 10^{62}~\rm erg$.  Its potential power output   
is $P_{rot}\approx 10^{44\rightarrow 47}~{\rm erg~s^{-1}}$ for spin-down timescales of $10^{10 \rightarrow 7}$ yr (Martini 2004).  
The power output, assuming a spin-down period of $10^8$ yr, is shown as the upper solid line labeled ``Pure Rotation" in Fig. 1.  The line lies well above the observations demonstrating that spin is a potentially important power source.

Lacking measurements of their
true accretion rates, we assume initially that all systems are accreting at the same
 rate $\dot{m}$ in units of the Eddington limit, and we evaluate their spin parameters, $j$, using the Nemmen et al. (2007) model.   
 Here the Eddington accretion rate is of the form $\dot M_{\rm Edd}=2.2 \epsilon^{-1}M_{\rm BH,9}~\rm M_\odot ~yr^{-1}$, where $\epsilon =0.1$ and 
 black hole mass is given in units of $10^9~\rm M_\odot$.  We further assume
 a viscosity parameter $\alpha = 0.2$, which is within the approximate range expected in a turbulent accretion flow (see Meier 2001, Nemmen et al. 2007). 
Varying $\alpha$ between $0.04-0.3$ in Nemmen's model does not change our conclusions significantly.

The solid line labeled ``$j=1$" in Fig. 1 shows the calculated jet power from maximally spinning holes as a function of  mass. 
The parallel broken lines show jet power calculated for lower spin parameters.  
In order to account for the high jet powers of MS0735 and Hercules A, the accretion rate must be
 $\dot m = 0.02$ or larger.  We therefore have adopted this accretion rate in Fig.~1.
Taken at face value, the data are consistent
with spin parameters lying in the broad range between
 $0.01 \lesssim j \lesssim 1$, and with a median value of $\simeq 0.6$.  However, due to the number of unknown variables, we are unable to place interesting constraints on the spin parameters of individual objects. 
 
The assumption of constant $\dot m$ implies a physical mass accretion rate $\dot M$ that increases in proportion to black hole mass.  With a black hole mass of about $6.4\times 10^{9} \rm ~M_{\odot}$ (Gebhardt \& Thomas 2009), M87 would then be accreting at $\dot M \sim 2.8~ \rm M_\odot ~yr^{-1}$.   Accretion at this rate would deplete its molecular gas reservoir in $3\times 10^6$ yr, which is inconsistent with the age of its current AGN outburst $7\times 10^7~\rm yr$ (Forman et al. 2007). 
Moreover, the gravitational binding energy released 
would dramatically exceed the observed mechanical power and radiation emerging from the nucleus,  unless the binding energy is being advected into the SMBH.  Other systems
suffer similar problems although their constraints are not as tight.
For these reasons, the assumption that all systems are accreting at or near $\dot{m}=0.02$ seems to be unrealistic.

 The distribution in Fig. 1 is equally consistent with all systems harboring SMBHs with high spin parameters but with broadly varying accretion rates, or with
both the accretion rate and spin varying widely.   We cannot
distinguish between these possibilities, except perhaps in the most powerful  systems.  

\subsection{A Constraint on $\dot m_{\rm crit}$}

Whether AGN release their energy in the form of radiation from a disk or in a jetted outflow is thought to depend on the the mass of the black hole and its accretion rate.   When the accretion rate approaches the Eddington limit,  AGN power emerges primarily as radiation from an optically thick,
geometrically thin disk.   When the accretion rate falls below a critical value in Eddington units, AGN power emerges in a jetted (radio) outflow (Narayan \& McClintock 2008).   Observations of Galactic X-ray binaries suggest $\dot m_{\rm crit} \sim 10^{-2}$ to $10^{-1}$ (Gallo, Fender, \& Pooley 2003,  Falcke, H., K{\"o}rding, E., \& Markoff, S.\ 2004,  
Churazov et al. 2005), but its precise value is unknown.     The objects in Fig. 1 are strong radio sources, yet they show little evidence for strong, unresolved nuclear ultraviolet emission
that is characteristic of accretion near the Eddington rate.  The accretion rates in Eddington units required to power their AGN are typically $10^{-4}$ and below, which is consistent with their being RIAFs or ADAFs.

MS0735 and Hercules A, with jet powers exceeding $10^{46}~\rm erg~s^{-1}$,  have the highest AGN power in our sample and are among the most powerful AGN known.
Yet, despite their quasar-like powers,  neither system shows
nuclear activity, such as bright optical and UV emission, that is normally associated with quasars (McNamara et al. 2009, Nulsen et al. 2005).  Therefore, their AGN
are unlikely to be accreting near the Eddington rate.  However, the interesting combination of extraordinarily high AGN power combined with such 
feeble nuclear luminosities suggest that their current accretion rates lie close to but below $\dot m_{\rm crit}$ (but see Sternberg \& Soker 2009 for a different point of view).
For $j\simeq 1$, the jet powers of MS0735 and Hercules A would imply $\dot m\simeq \dot m_{\rm crit} \simeq 0.02$, 
which is consistent with theoretical and observational estimates  of $\dot m_{\rm crit}$ from X-ray binaries (Narayan \& McClintock 2008, Gallo, Fender, \& Pooley 2003,  Falcke, H., K{\"o}rding, E., \& Markoff, S.\ 2004).  This value of $\dot m_{\rm crit}$ corresponds to physical accretion rates
of roughly $2.2~ \rm M_\odot ~yr^{-1}$ and  $1.1 ~\rm M_\odot ~yr^{-1}$, respectively.  Slightly larger values 
of $\dot m_{\rm crit}$ are found by assuming their AGN are powered by accretion onto a Schwarzschild black hole (Churazov et al. 2005).
In order to maintain MS0735's jet power while easing back on MS0735's spin parameter would require boosting its accretion
rate to a level that is alarmingly high relative to its gas supply
(see McNamara et al. 2009).   Thus, $\dot m_{\rm crit}$  implied by these systems cannot be much larger.  Alternatively, higher jet power can be achieved with a lower spin parameter, if their SMBHs are considerably more massive than the $M_{\rm BH}-L_{\rm bulge}$ relation predicts (see Lauer et al. 2007 and McNamara et al. 2009 for discussions).  
Apart from these caveats, our analysis implies that the transition from low to high radiative efficiency occurs at accretion rates of no less than a few percent.

\section{Bondi Accretion}

Ignoring spin power for the moment,  the accretion rate required to power the
AGN in our sample assuming $P_{\rm jet} = \epsilon \dot {M}c^2$ is shown on the right hand side of Fig. 1.  
The value of $\epsilon$ in any given system is poorly known and can vary between  $0.06$ and $0.42$ depending on whether we are dealing with respectively, a non-rotating black hole or a maximally-spinning black hole.   Given this uncertainty, we have adopted $\epsilon =0.1$ for all objects, which has become standard practice in the field.
The accretion rates vary from $\dot M \simeq 10^{-4} \msunyr$ in the gE galaxy M84 to several $\msunyr$ in Hercules A and MS0735.   
The Bondi accretion rate from a hot atmosphere scales as 
$\dot M_{\rm B} \propto n_e (kT)^{-3/2} M_{BH}^2$, where $n_e$ and $T$ are, respectively, the electron density and gas temperature at the Bondi radius, and $M_{BH}$ is the black hole mass.
Bondi accretion can  be effective only when the hot atmosphere is sufficiently dense near the Bondi radius to feed the SMBH at a rate consistent with $P_{\rm jet}$. 

In Fig. 2 we
plot the ratio of jet power to Bondi accretion power against estimated black hole mass. We assume  $P=\eta \epsilon \dot Mc^2$ where the efficiency of accretion through the Bondi sphere is $\eta = 1$.  In other words, all of the mass reaching the Bondi
radius is assumed to be accreted onto the SMBH.   Fig. 2 shows  that with the exception of  the lower power systems residing within or near the shaded region of the plot, Bondi accretion would have great difficulty powering cluster AGN outbursts. Rafferty et al. (2006), using the data presented here, and Hardcastle et al. (2007), using powerful
radio galaxies, reached similar conclusions.

Falling gas temperatures and rising gas densities near the (unresolved) Bondi radii, in addition to the possibility
that the black hole masses may be larger than their bulge luminosities imply, would increase the number of objects lying within the shaded region in
Fig. 2 (see Rafferty et al. 2006 for a thorough discussion).   However, this effect will be offset to a degree by the overly optimistic assumption that $\eta =1$.
Mass lost to winds blowing from the accretion disk and the need to shed angular momentum from the accreting gas
(eg., Neilsen \& Lee 2009, Proga 2009, Soker et al. 2009) are expected to drive $\eta$ well below unity (Allen et al. 2006, Merloni \& Heinz 2008, Benson \& Babul 2009, Li \& Cao 2010).  For example, Allen et al. (2006), Merloni \& Heinz (2008) found that only a few percent of the matter reaching the Bondi radius is
actually converted into jet power, which is consistent with $\eta < 1$.  Accretion at this level would be able power low luminosity AGN found in elliptical galaxies (Allen et al. 2006).  However, it strengthens our conclusion that Bondi accretion from the hot atmosphere alone probably cannot fuel the most powerful AGN in clusters.  

\section{Cold Accretion}

In Fig. 3 we plot {\it total} molecular gas mass against jet power for the objects with gas mass measurements
available in the literature. The molecular gas masses generally lie between  $10^{9}  M_{\odot}$ to  $10^{11}~M_{\odot}$.   Only M87's upper limit of less than $8 \times 10^6~M_\odot$ (Tan et al. 2008) lies substantially below this range.   Gas masses were corrected to our adopted cosmology when necessary; details of the gas mass analysis
can be found in the references given in the caption to Fig. 3.
The gas masses needed to fuel the AGN, $M_{acc} \sim E_{cav}/ 0.1c^{2}=10^{6}~M_{\odot}-10^{9}~M_{\odot}$, lie well within the observed range seen in Fig. 3.  If these AGN are powered primarily by accretion of molecular gas,  a correlation between the gas supply 
and jet power would be expected.  Spin models requiring high accretion 
rates  would also predict a correlation.  
Yet no correlation is seen between molecular gas mass and jet power {\sl within the range of jet power shown in Fig. 3.} The Spearman rank order correlation coefficients for the sample including and excluding upper limits are 0.35 and 0.60, respectively.  These statistical figures of merit
confirm the absence of an evident correlation in Fig. 3.  In fact, for a given jet power the molecular gas reservoirs vary
in mass by more than two orders of magnitude.  Furthermore, the large gas supply relative to AGN power in most systems suggests that very little molecular gas is currently reaching 
the black hole.  

The large scatter and absence of a correlation may be related to several factors.  The most important factor may be the presence or absence of star formation, which we discuss
further below.  The high star formation rates in many of the objects in our sample suggests that most of their molecular gas is being consumed by star formation (Rafferty et al. 2006, O'Dea et al. 2008).   Some of the gas that is not consumed by star formation may be driven away from the nucleus by AGN (Sternberg \& Soker 2009).  In addition, temporal variations in accretion rate related perhaps to dynamical interactions with neighboring galaxies may also be contributing to the scatter.  Finally, because in most systems only a small fraction of the molecular gas mass is required to fuel the AGN, a
real correlation may be obscured by these and other factors.  Whatever the important factors may be, AGN
power seems to be largely decoupled from the total molecular gas supply. 

\subsection{Accretion Efficiency per AGN Outburst}

Another way to look at this problem is to evaluate the fraction of the gas mass that must be consumed by the AGN in order to power it. 
As a point of reference, the average mass accretion efficiency of SMBHs implied by scaling relations between bulge and black hole mass
is $\simeq 1.4\times 10^{-3}$ (Magorrian et al. 1998, H\"aring \& Rix 2004).   In other words,  for each unit of mass that fell into nuclear black holes, roughly 700 units of mass formed  stars.
Assuming that most of the molecular gas in BCGs is consumed by star formation (O'Dea et al. 2008), we can define the accretion efficiency per AGN outburst as $E_{\rm jet}/0.1M_{\rm mol}c^2$.  
Using this definition, a plot of accretion efficiency versus molecular gas mass  (Fig. 4)  shows a median value of  
approximately $6\times 10^{-4}$, a value that lies below but is roughly consistent with the fraction of gas that is expected to be consumed
by the black hole based on scaling relations (H\"aring \& Rix 2004).  However, the accretion efficiency at a given gas mass varies by more than three orders of magnitude.
So it is unclear whether the similarity of these two figures is more than a coincidence. 

The extremes are represented by Abell 1068, a cluster with a burgeoning central galaxy hosting a relatively weak AGN, and MS0735, whose dormant, gas-poor BCG hosts an extraordinarily powerful AGN, but no appreciable star formation.  Abell 1068's BGG contains $\sim 10^{11}~M_\odot$ of molecular gas (Edge 2001) and is experiencing star formation at a rate of $\sim 60 ~M_\odot ~\rm yr^{-1}$ (McNamara et al. 2004).   The BCG has a weak radio AGN indicating a current accretion efficiency below $10^{-5}$. 
It is a strong far infrared source (Edge et al. 2010), some of which may be associated with a buried AGN  (Quillen et al. 2008).  The energy being
released in the infrared may indicate a somewhat  higher accretion efficiency  than its radio power indicates, but its mechanical energy clearly falls well below the sample mean.   

Most of the objects
in this sample with molecular gas masses above a few $10^9~M_\odot$ are experiencing star formation and thus bulge growth at some level.  All of the objects
with gas masses above  $10^{10}~M_\odot$, located to the upper left of Fig. 4,  are experiencing star formation above several tens of solar masses per year (O'Dea 2008).   By virtue of their AGN outbursts, all are experiencing  black hole growth.  Rafferty et al. (2006) found
many cooling flow BCGs growing parallel to 
the slope of the black hole versus bulge mass scaling relations, which is complementary to our finding that the median accretion efficiency here lies close to the expected value from black
hole scaling relations.  Abell 1068 is an outlier in the sense that its black hole appears to be growing more slowly than expected for its current star formation rate.  

With an accretion efficiency of approximately 10 per cent, MS0735 represents the opposite extreme.  A search for molecular gas in MS0735 in the
CO[$1\rightarrow 0$] emission line by Salom\`e \& Combes (2008) revealed an upper limit of $\sim 10^{10}\ ~M_\odot $.   An HST image of the BCG has revealed
weak nuclear dust features,  and its star formation rate, based on both far ultraviolet
(McNamara et al. 2009) and mid infrared observations (Donahue et al. 2010, in preparation), lies below $0.25 ~M_\odot ~\rm yr^{-1}$. 
 These properties taken together are consistent with other BCGs with gas reservoirs
of order $10^9  ~M_\odot $ or less. 
The implication is that the ratio of MS0735's estimated molecular gas mass to its total jet power indicates that a substantial fraction of its existing gas
supply must have been consumed by its SMBH in the past $10^8$ yr.  If true, the gas must have efficiently  shed its angular momentum and it must
have done so  without forming an
appreciable number of stars, which would be difficult to understand (see McNamara et al. 2009 for a discussion). 
Hydra A, MKW3S, and Cygnus A are
similar but less extreme.   In fact, these objects appear to congregate together at the high efficiency and relatively low gas mass quadrant to the lower right of Fig. 4.  Given their
relatively modest gas masses, it is surprising that they they are among the most powerful AGN known in clusters.  

Oca{\~n}a Flaquer et al. (2010) recently noted
relatively modest molecular gas masses of a few $10^8~M_\odot$ in powerful, nearby radio galaxies.  In addition,
Emonts et al. (2007) reported an anti-correlation between radio source size, a property that is related to jet power, and neutral hydrogen mass in giant elliptical galaxies.
These results suggest that  the trend we see in BCGs may be related to a more general phenomenon of high jet powers associated with relatively gas-poor host galaxies.

Being the most powerful FR II radio source in the nearby Universe, Cygnus A is worth a brief discussion. An analysis of the shock front in the X-ray halo surrounding its radio lobes gives  Mach number 1.4,  total shock energy $E_s=2\times 10^{60}~\rm erg$,
and a mean jet power of $4\times 10^{45}\rm ~ erg~s^{-1}$ (Nulsen et al. 2010, in preparation).  This power measurement is
substantially less than that of Wilson et al. (2006),  which treated the shock as very strong.  Powering Cygnus A by accretion would have consumed  more than $\sim 10^7M_\odot$ of gas over the past $1.6\times 10^7$ yr.  For comparison, Salome \& Combes (2003) found  less than $1.5\times 10^9M_\odot$ of molecular gas in its bulge, which is more than enough gas to feed the outburst.   Nevertheless, its accretion efficiency, which exceeds $7\times 10^{-3}$,  indicates that it is fueling its AGN relatively efficiently (see Fig. 4.)

The outburst energies of Abell 1835 and Zw 3146, both of which are indicated in Fig. 4,
are comparable to that of Cygnus A.  However, their accretion efficiencies are an order of magnitude or so lower, placing them to the upper left in Fig. 4.   Unlike Cygnus A, their bulges contain $\sim 10^{11}~M_{\odot}$ of molecular gas and they are forming stars at rates approaching $100~M_\odot~\rm yr^{-1}$.  Their locations in Fig. 4 imply that 
gas-rich systems with high star formation rates are unable to channel fuel onto the nucleus as efficiently as gas-poor systems.  Star formation associate with gas stalled
in circumnuclear disks may be inhibiting the flow of gas onto the AGN. 

Assuming all AGN in this sample are powered by accretion,  the large variation in accretion efficiency with respect to gas mass seen in Fig. 4 must be contributing
to the variation in jet power with respect to gas mass seen in Fig. 3.   Some of the variation in accretion efficiency may be related to the value of the
mass to energy conversion efficiency, $\epsilon$, which depends on the spin of the black hole.
$\epsilon$ can vary by a factor of 7 depending on whether the black hole is a non-rotating Schwarzschild black hole or a maximally spinning Kerr black hole.
The scatter in Fig. 4 is much larger than this factor, so the spin of the black hole alone cannot account for it.

In light of this discussion,  it is unclear whether spin powers all radio AGN or any AGN, for that matter.  Nevertheless, those systems with high (apparent) accretion efficiencies and small
gas reservoirs  that would have difficulty powering their AGN by accretion alone may be the best candidates  (see also Paggi et al. 2009 and Hart et al. 2009).  The objects meeting
this criterion our sample: Cygnus A, Hydra A, MS0735, MKW3s, are located to right side of Fig. 4.   If their current AGN outbursts are powered by accretion, then their black
holes apparently consumed between one part in a few to one part in one hundred of their entire gas supply, possibly indicating a highly efficient mode of accretion.   While it would be tempting to attribute powerful AGN residing in gas poor galaxies to
spin, current BZ-based spin models (e.g., Nemmen et al. 2007, Benson \& Babul 2009) also require relatively high
accretion rates.  Therefore,  a mechanism that is able to tap spin power efficiently at relatively low accretion rates, i.e.,  $P_{jet} > \dot Mc^2$,  may be needed to explain these systems, 
particularly the AGN in MS0735. 
We note that existing BZ models still fail to make $P_{jet} > \dot Mc^2$.

 \section {Concluding Remarks}

We have examined two possible scenarios for powering AGN in brightest cluster galaxies: accretion of cold molecular gas and black hole spin.  Understanding how radio AGN are powered is central to many questions related to the evolution galaxies and black holes,  including how AGN feedback operates at late times (eg., Croton et al. 2006, Somerville et al. 2008). Although we are generally unable to distinguish
between these two related scenarios, we find that AGN power in BCGs would be consistent with both a broad range of spin parameter and accretion rate.  We find no significant correlation between
jet power and molecular gas mass in these systems. We have identified several powerful AGN associated with BCGs that have surprisingly low gas reserves.  
Although accretion of cold gas  must be important at some level,  this study shows that AGN power is poorly correlated with the {\it total} molecular gas mass of the
the host BCG.  Molecular gas  is clearly associated with star formation (O'Dea et al. 2008), so perhaps most of the gas is being consumed by star formation before it is
able to flow into the nucleus.
Bondi accretion from  hot atmospheres, which has become a staple in AGN feedback models,  may be able to fuel weak AGN (Allen et al. 2006), but would have great difficulty supplying enough gas to power the most energetic AGN.  

 If radio AGN are powered instead by black hole spin, the observed distribution of jet power implies a broad range of spin parameter,  given the spin model adopted here
 (Nemmen et al. 2007).
BZ-based models such as Nemmen's  require substantial accretion rates in order to access spin power.  So unfortunately it is impossible to place interesting constraints on the spin parameter  based on jet power alone.  We have highlighted several powerful AGN residing in relatively gas-poor bulges, which  we suggest as good candidates for jet  powering by black hole spin.  However, they may have difficulty achieving their power output even with spin parameters approaching
unity, unless their spin is tapped with high efficiency, i.e.,  $P_{jet} >  Mc^2$ (eg., Garofalo et al. 2010), or their black holes are more massive than expected.  

Our understanding of the accretion process in AGN will advance significantly in the future when ALMA becomes operational, and we are able to disentangle
the molecular gas fueling star formation from the nuclear gas that will eventually plunge into the
black hole.  Likewise, future large aperture X-ray and optical/IR telescopes with sub arcsecond resolution are needed to explore the environments of AGN nearer to their Bondi 
spheroids.

\acknowledgments
We thank M. Ruszkowski, D. Evans, and A. Babul for enlightening conversations, and we thank the referee
for offering suggestions that improved the paper.  This research
was supported in part by Chandra Guest Observer grant G09-0140X, and a generous grant from the Natural Sciences and Engineering Research
Council of Canada.

\clearpage



\begin{figure}
\epsscale{0.8}
\plotone{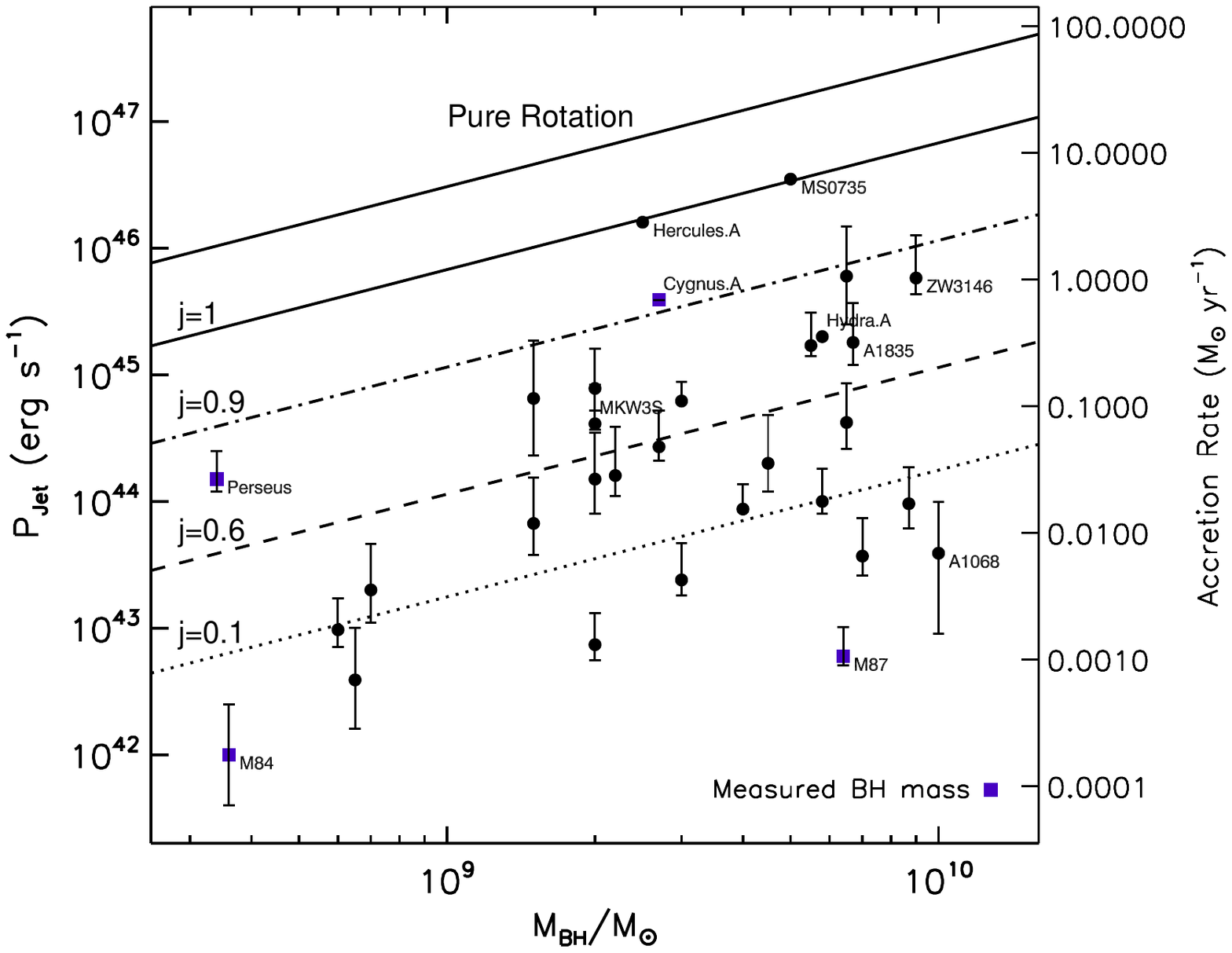}
\caption{The jet (cavity) power versus black hole mass. The top line labeled ``Pure Rotation" represents the classical energy available from a maximally rotating black hole. The remaining lines represent the jet power of hybrid model from critically accreting black holes, $\dot{m}_{c}=0.02$, with spin parameters between 1 to 0.1.  The filled blue points represent jet powers from central cluster galaxies with measured black hole masses (Wilman et al. 2005; Gebhardt \& Thomas 2009).
    Cavity and shock front data are from Rafferty et al. (2006), Wise et al. (2007), Nulsen et al. (2005), and McNamara et al. (2009). The right axis shows the accretion
    rate that would be required to power the AGN outburst ignoring spin and assuming $P_{\rm jet}=0.1 \dot Mc^2$.}
\end{figure}

\clearpage

\begin{figure}
\epsscale{0.95}
\plotone{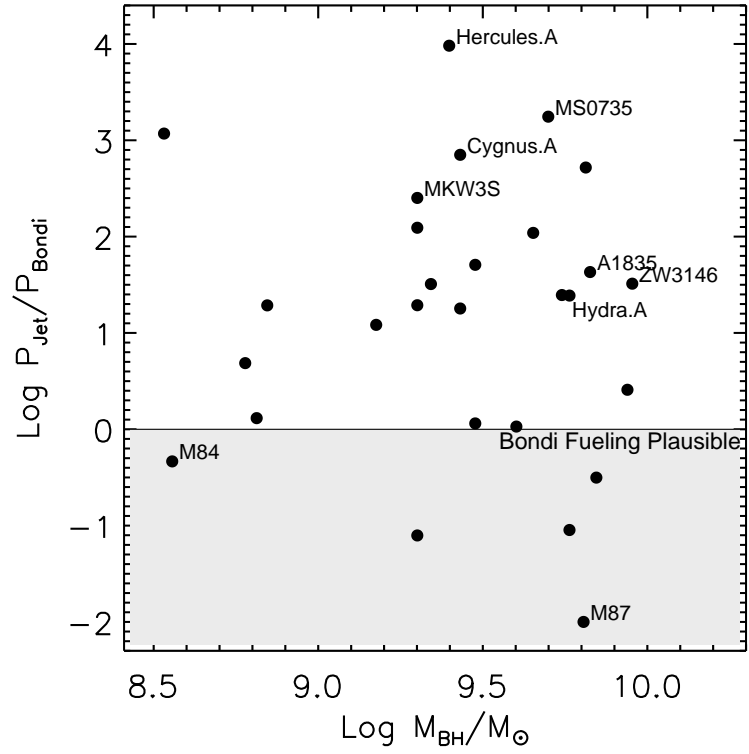}
\caption{The ratio of jet power, $P_{jet}$, to Bondi accretion power, $P_{Bondi}=0.1\dot{M}_{B}c^{2}$, for 28 objects against their black hole mass taken 
    from Rafferty et al. (2006).}
\end{figure}

\clearpage
\begin{figure}
\epsscale{0.95}
\plotone{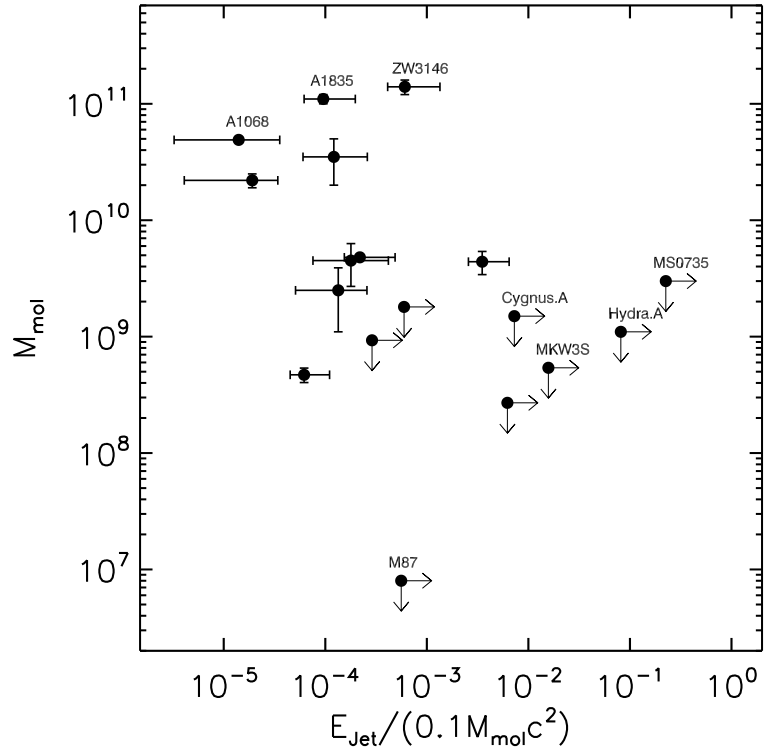}
\caption{Molecular gas masses from Edge et al. (2001), Salome and Combes (2003, 2004, 2006), Salome et al. (2008), and Tan et al. (2008)
    versus jet power.  Upper limits are $3\sigma$ values. }
\end{figure}

\clearpage
\clearpage
\begin{figure}
\epsscale{0.95}
\plotone{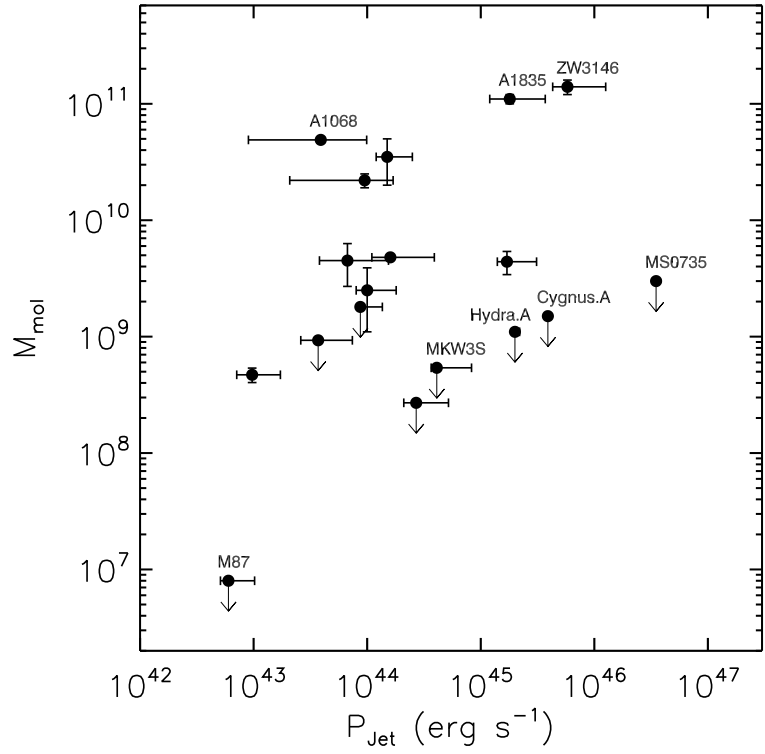}
\caption{Molecular gas mass vs accretion efficiency per AGN outburst.  The high efficiency objects on the right side of
    the plot are powerful AGN located in relatively gas-poor host galaxies.}
\end{figure}

\clearpage
\clearpage
\def\arraystretch{1.6}
\begin{deluxetable}{cccccc}
\tabletypesize{\scriptsize}
\tablewidth{0pt}
\tablecaption{Sample Data}
\tablehead{ 
   \colhead{} & \colhead{$P_{jet}$} & \colhead{$M_{BH}$} & \colhead{$M_{mol}$} & \colhead{Efficiency} &  \colhead{} \\
   \colhead{Object} & \colhead{($10^{42} ~\rm erg~s^{-1}$)} & \colhead{($10^{9}$ $M_{\odot}$)} & \colhead{($10^8$ $M_{\odot}$)} & \colhead{  ($10^{-4}$)  } & \colhead{References} }

\startdata
   A85 &  $37_{-11}^{+37}$ &  7.0 & $<9.3$ & $>3.0$ & 5,9 \\ 
   A133 &  $620_{-20}^{+260}$ &  3.0 & ... & ... & 9 \\ 
   A262 &  $9.7^{+7.5}_{-2.6}$ &  0.6 & $4.7 \pm 0.67$ &  ${0.6}_{-0.2}^{+0.5}$ & 5,9 \\ 
   Perseus &  $150^{+100}_{-30}$ &  0.34 & $350 \pm 150$ &  ${1.2}_{-0.6}^{+1.4}$ & 7,9,11 \\ 
   2A0335+096 &  $24^{+23}_{-6}$ &  3.0 & ... & ... & 9 \\ 
   A478 &  $100^{+80}_{-20}$ &  5.8 & $25\pm 14$ &  ${1.3}_{-0.8}^{+1.2}$ & 1,9 \\ 
   MS 0735.6+7421 &  $35000$ &  5.0 & $<30$ & $>2000$ & 3,8,9 \\ 
   PKS 0745-191 &  $1700^{+1400}_{-300}$ &  5.5 & $44 \pm 9.9$ &  ${35}_{-9.3}^{+29}$ & 5,9 \\ 
   4C 55.16 &  $420^{+440}_{-160}$ &  6.5 & ... & ... & 9 \\ 
   Hydra A &  $2000^{+50}_{-50}$ &  5.8 & $<11$ & $>1000$ & 1,9,12 \\ 
   Zw 2701 &  $6000^{+8900}_{-3500}$ &  6.5 & ... & ... & 9 \\
   Zw 3146 &  $5800^{+6800}_{-1500}$ &  9.0 & $1400 \pm 200$ &  ${6.0}_{-1.9}^{+7.4}$ & 1,9 \\
   A1068 &  $39^{+60}_{-60}$ &  10 & $490 \pm 30$ &  ${0.1}_{-0.2}^{+0.2}$ & 1,9 \\
   M84 &  $1.0^{+1.5}_{-0.6}$ &  0.36 & ... & ... & 9 \\
   M87 &  $6.0^{+4.2}_{-0.9}$ &  6.4 & $<0.08$ & $>5.0$ & 2,9,10 \\
   Centaurus &  $7.4^{+5.8}_{-1.8}$ &  2.0 & ... & ... & 9 \\
   HCG 62 &  $3.9^{+6.1}_{-2.3}$ &  0.65 & ... & ... & 9 \\
   A1664  &  $95.2^{+74}_{-74}$  &... &  $220 \pm 30$& $0.2^{ +0.15}_{-0.15}$& 13\\

   A1795 &  $160^{+230}_{-50}$ &  2.2 &  $48 \pm 6$&  ${2.0}_{-0.6}^{+2.7}$ & 6,9 \\
   
   A1835 &  $1800^{+1900}_{-600}$ &  6.7 & $1100 \pm 100$ &  ${0.9}_{-0.3}^{+1.0}$ & 1,9 \\
   PKS 1404-267 &  $20^{+26}_{-9}$ &  0.7 & ... & ... & 9 \\
   A2029 &  $87^{+49}_{-4}$ &  4.0 & $<18$ & $>6.0$ & 5,9 \\
   A2052 &  $150^{+200}_{-7}$ &  2.0 & ... & ... & 9 \\
   MKW 3S &  $410^{+420}_{-44}$ &  2.0 & $<5.4$ & $>200$ & 5,9 \\
   A2199 &  $270^{+250}_{-60}$ &  2.7 & $<2.7$ & $>60$ & 5,9 \\
   Hercules A &  $16000$ &  2.5 & ... & ... & 4,9 \\
   3C 388 &  $200^{+280}_{-80}$ &  4.5 & ... & ... & 9 \\
   3C 401 &  $650^{+1200}_{-420}$ &  1.5 & ... &...  & 9 \\
   Cygnus A &  $3900$ &  2.7 & $<15$ & $>70$ & 5,9 \\
   Sersic 159/03 &  $780^{+820}_{-260}$ &  2.0 & ... & ... & 9 \\
   A2597 &  $67^{+87}_{-29}$ &  1.5 & $45 \pm 18$ &  ${1.7}_{-1.0}^{+2.4}$ & 1,9 \\
   A4059 &  $96^{+89}_{-35}$ &  8.7 & ... & ... & 9 \\ 
\enddata
\tablerefs{(1) Edge et al.\ 2001; (2) Gebhardt et al.\ 2009; (3) McNamara et al.\ 2009; (4) Nulsen et al.\ 2005; (5) Salome et al.\ 2003; (6) Salome et al.\ 2004; (7) Salome et al.\ 2006; (8) Salome et al.\ 2008; (9) Rafferty et al.\ 2006; (10) Tan et al.\ 2008; (11) Wilman et al.\ 2005; (12) Wise et al.\ 2007; (13) Kirkpatrick et al. 2009}
\end{deluxetable}

\end{document}